\documentclass[3p,times]{elsarticle}
\pdfoutput=1

\usepackage{ecrc}

\volume{00}

\firstpage{1}

\journalname{Nuclear Physics A}

\runauth{A. Kesich et al.}

\jid{nupha}

\usepackage{amssymb}
 \usepackage{lineno}

\usepackage[figuresright]{rotating}

	\newcommand{\bbbar}{b$\bar{\textrm{b}}$ }
	\newcommand{\RAA}{$R_{AA}$}
\begin{document}

\begin{frontmatter}

%% Title, authors and addresses

%% use the tnoteref command within \title for footnotes;
%% use the tnotetext command for the associated footnote;
%% use the fnref command within \author or \address for footnotes;
%% use the fntext command for the associated footnote;
%% use the corref command within \author for corresponding author footnotes;
%% use the cortext command for the associated footnote;
%% use the ead command for the email address,
%% and the form \ead[url] for the home page:
%%
%% \title{Title\tnoteref{label1}}
%% \tnotetext[label1]{}
%% \author{Name\corref{cor1}\fnref{label2}}
%% \ead{email address}
%% \ead[url]{home page}
%% \fntext[label2]{}
%% \cortext[cor1]{}
%% \address{Address\fnref{label3}}
%% \fntext[label3]{}

\dochead{}
%% Use \dochead if there is an article header, e.g. \dochead{Short communication}

\title{Measurements of $\Upsilon$ Production and Nuclear Modification Factor at STAR}

%% use optional labels to link authors explicitly to addresses:
%% \author[label1,label2]{<author name>}
%% \address[label1]{<address>}
%% \address[label2]{<address>}

\author{Anthony Kesich for the STAR Collaboration}

\address{One Shields Ave., Davis, CA, 95616, USA}

\begin{abstract}
Thermal suppression of quarkonium production in heavy-ion collisions, due to Debye screening of the quark-antiquark potential, has been proposed as a clear signature of Quark-Gluon Plasma (QGP) formation. At RHIC energies, the $\Upsilon$ meson is a clean probe of the early system due to negligible levels of enhancement from \bbbar recombination and non-thermal suppression from co-mover absorption. We report on our measurement of the $\Upsilon \to e^+ e^-$ cross section in Au+Au collisions at $\sqrt{s_{NN}}$=200 GeV. We compute the Nuclear Modification Factor by comparing these results to new p+p measurements from 2009 (21 pb$^{-1}$ in 2009 compared to 7.9 pb$^{-1}$ in 2006). In order to have a complete assessment of both hot and cold nuclear matter effects on Upsilon production we also report on results from d+Au collisions.
\end{abstract}

\begin{keyword}
%% keywords here, in the form: keyword \sep keyword
STAR \sep QGP \sep Quarkonium \sep Upsilons
%% MSC codes here, in the form: \MSC code \sep code
%% or \MSC[2008] code \sep code (2000 is the default)

\end{keyword}

\end{frontmatter}

%%
%% Start line numbering here if you want
%%
%\linenumbers

%% main text
\section{Introduction}
Understanding the production and evolution of the Quark-Gluon Plasma is one of the key goals of the RHIC program. By characterizing the QGP, we can achieve a better understanding of the properties of the strong force and nuclear matter. Heavy quarkonia are unique probes to help study this question. Being massive, heavy quarks are produced in initial hard scatterings and therefore carry information pertaining to the early state of the system. Furthermore, lattice-based calculations suggest that formation of a QGP should screen the inter-quark potential in heavy quarkonia and effectively melt these mesons giving us a direct observable of the temperature of the system \cite{Matsui1986416, Strickland:2011kx}.

\section{Analysis}
The analyses shown herein were done using three different datasets. The d+Au dataset was collected in 2008, the p+p dataset in 2009, and the Au+Au dataset in 2010. All three were at $\sqrt{s_{NN}} = 200$ GeV. All three datasets were collected with a High Tower  trigger ($E_T\gtrsim4.3$ GeV in a single calorimeter tower). Furthermore, the p+p and d+Au datasets were also required to satisfy a dedicated Upsilon trigger. This trigger used energy deposited in the calorimeter to reconstruct candidate particles assuming any large deposits of energy came from electrons. If the reconstructed particle has a mass near the Upsilon, it is kept for further study. This trigger was not used for the 2010 Au+Au dataset thanks to increased data-taking bandwidth in STAR.

There were two main detectors used in these analyses: the Time Projection Chamber (TPC) and the Barrel Electromagnetic Calorimeter (BEMC). The TPC is a gas ionization detector used to track charged particles and measure their momentum and charge. Furthermore, by looking at the ionization rate of a track, we can identify particles via their characteristic energy loss. The BEMC is used to measure the energy of photons and electrons. When combined with momentum measurements from the TPC, this allows for electron identification via the ratio of Energy seen in the BEMC to the momentum measured in the TPC. Furthermore, the BEMC is a fast detector and serves as the triggering mechanism for both the High Tower and Upsilon triggers.

We reconstruct our Upsilon candidates by looking at unlike-sign (oppositely charged) pairs of identified electrons. Our total unlike-sign signal is composed of four main parts: a combinatorial background, Drell-Yan production, correlated \bbbar pairs, and $\Upsilon$(1S+2S+3S)$\to e^+ e^-$. The combinatorial background is modeled by looking at like-sign (same charge) pairs of identified electrons. The shape of the Drell-Yan signal is obtained from PYTHIA simulations. The \bbbar background is obtained from pQCD calculation by R. Vogt. The shape of our Upsilon signal comes mainly from detector effects such as bremsstrahlung radiation and momentum resolution. It is obtained via embedding PYTHIA simulations into our detector model and real data.

\section{Upsilon Production in p+p Collisions}

\begin{figure}[h]
\begin{center}
\includegraphics[width=200px]{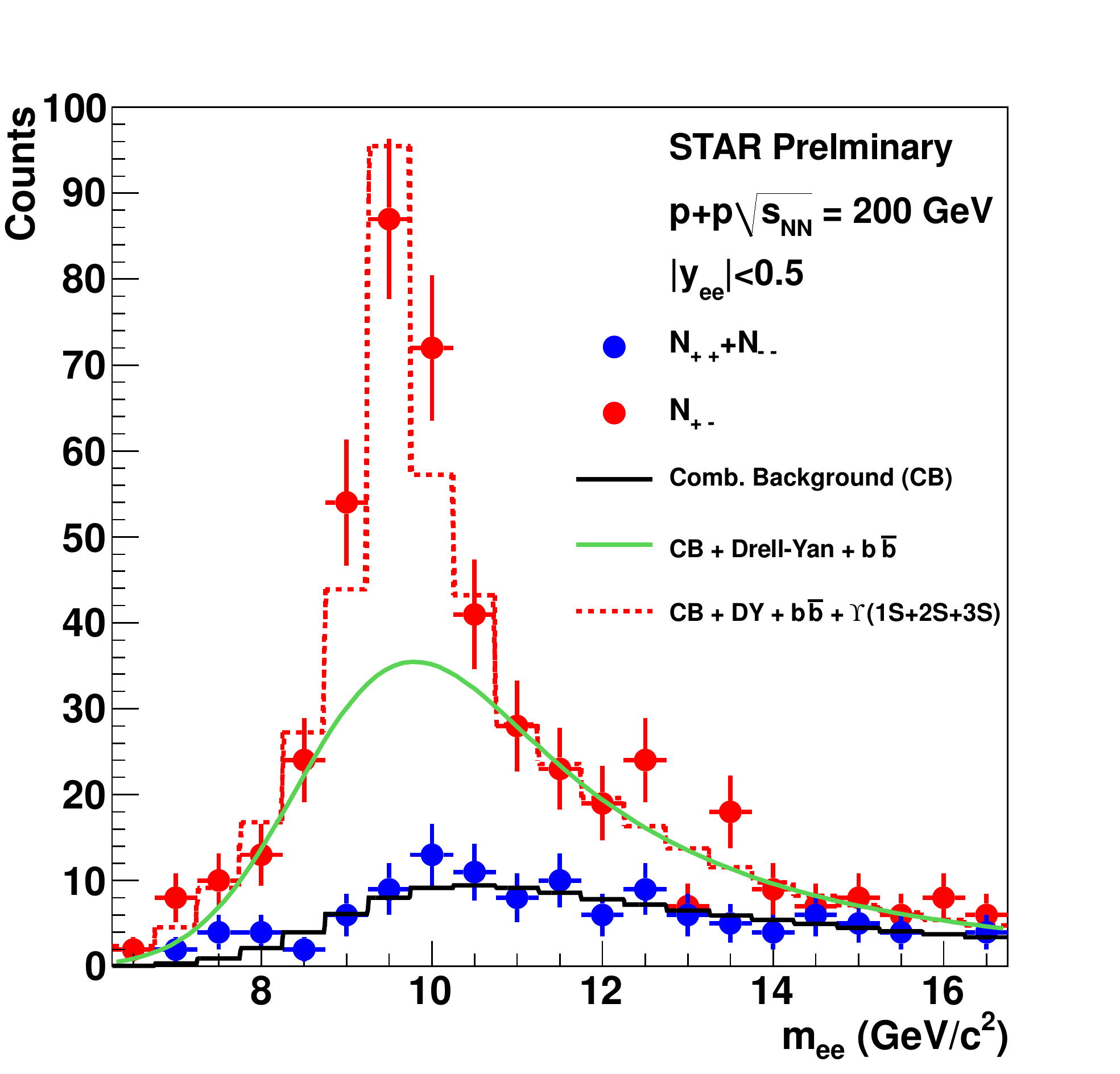}
\includegraphics[width=200px]{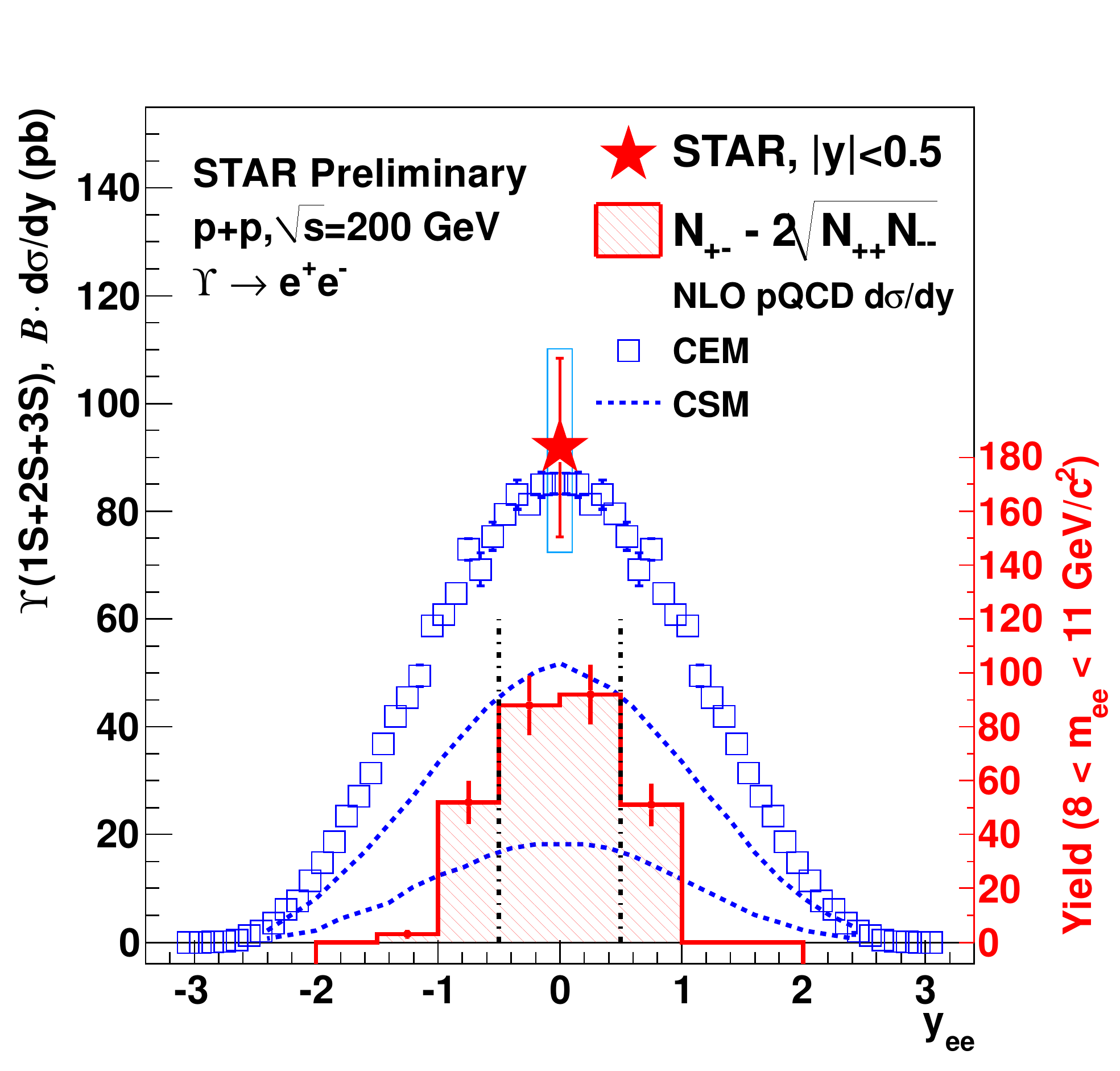}
\caption{\textit{Left:} Invariant mass spectrum of pairs of identified electrons. The red points are unlike-sign pairs, the blue are like-sign. The green curve represents the Drell-Yan and \bbbar background signals. The peaked shape of the background is due to kinematic constrains imposed by our Upsilon trigger. \textit{Right:} Comparison of our measurement to theory. The dotted blue line is a Color Evaporation Model calculation \cite{Brodsky:2010uq}. The blue squares are calculated using a Color Singlet Model \cite{Frawley2008125}.}
\label{fig:pp2009_ups}
\end{center}
\end{figure}

In the left plot of Fig. \ref{fig:pp2009_ups}, the invariant mass spectrum for electron pairs is shown. The mass spectrum has contributions from an obvious kinematic peak (not the $\Upsilon$ peak) around 10 GeV which can be seen in both the like-sign and unlike-sign spectra. This peak is imposed by the dedicated Upsilon trigger due to geometrical and energetic constraints imposed on candidate events. In order to constrain this peak, we employ a simultaneous fitting procedure which fits to both the unlike-sign and like-sign signals at the same time. Since the kinematic constraints will be imposed on both, this gives us a stronger handle on our background shape than if we used the background alone. The Drell-Yan and \bbbar signals mentioned above are shown in green.  Our upsilons signal is modeled as the sum of three Crystal Ball functions (one for each state). We fit the shape of the Crystal Ball functions to embedding simulations and the relative ratios of the three states are fixed to PDG values. We then fit to the integral of our total signal (red dotted line).

When computing the yield of Upsilons, the line shapes are only used to facilitate the fitting process. Once we have a fit, we subtract the fitted yield of Drell-Yan, \bbbar, and the combinatoric background (green curve) from the raw unlike-sign data (red points). This helps remove any dependence the yield would have on the signal shape. Ultimately, we find a total of 145$\pm$26 $\Upsilon$ candidates. After applying efficiency corrections and accounting for the sampled luminosity, this we obtained a cross section of $\Upsilon$(1S+2S+3S)$\to e^+ e^-$ for $|y|< 0.5$ of 91.8$\pm$16.6$\pm$19 pb. This is a great statistical enhancement over our previous measurement of $\Upsilon$ production p+p collisions from 2006 of $114\pm38^{+23}_{-24}$ pb\cite{Collaboration:2010uq}.

The right plot of Fig. \ref{fig:pp2009_ups} shows a comparison between our measured cross section and results from theoretical calculations. The blue squares are a prediction using a Color Evaporation Model \cite{Frawley2008125} whereas the dotted blue line is computed using a Color Singlet Model \cite{Brodsky:2010uq}. Our measurement is clearly more consistent with CEM than CSM.

\section{Upsilon Production in d+Au Collisions}
The 2008 d+Au dataset was collected in a similar manner to the p+p dataset. We again used the dedicated Upsilon trigger to collect the data. However, to account for the combinatoric background in this case, we subtracted the geometric mean of our like-sign pairs from the unlike-sign pairs. The resulting signal then only included our physical (i.e. non-combinatorial) signals: Drell-Yan, \bbbar coalescence, and $\Upsilon$(1S+2S+3S). We measure a combined cross section for all three processes in the mass range 7 GeV$<$m$<$11 GeV of $35\pm4\pm5$ nb for $|y|<0.5$.

\begin{figure}[h]
\begin{center}
\includegraphics[width=200px]{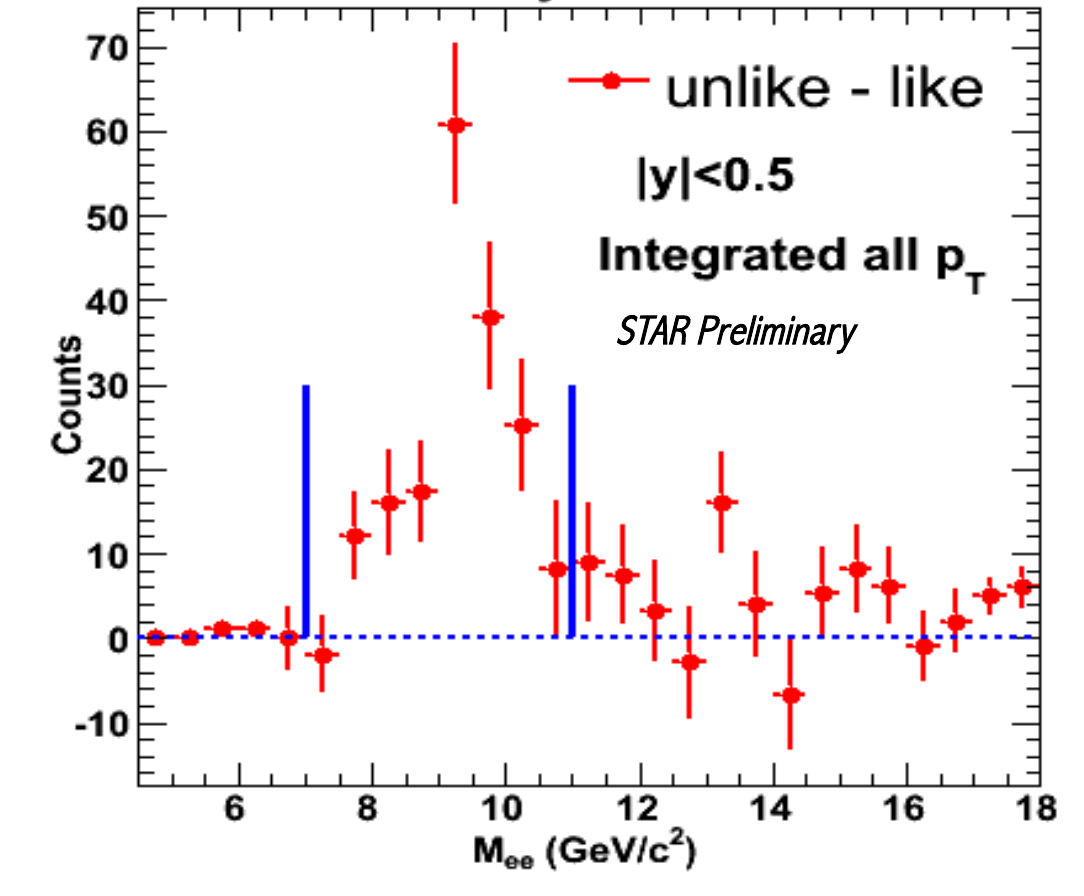}
\caption{Invariant mass spectrum of unlike-sign minus like-sign pairs of identified electrons. This contains the Upsilon signal, Drell-Yan production, and \bbbar recombination. We see an $8\sigma$ signal in the mass window of 7-11 GeV.}
\label{fig:dAu2008_ups}
\end{center}
\end{figure}

\section{Upsilon Production and Nuclear Modification Factor in Au+Au}
\begin{figure}[h]
\begin{center}
\includegraphics[width=152px]{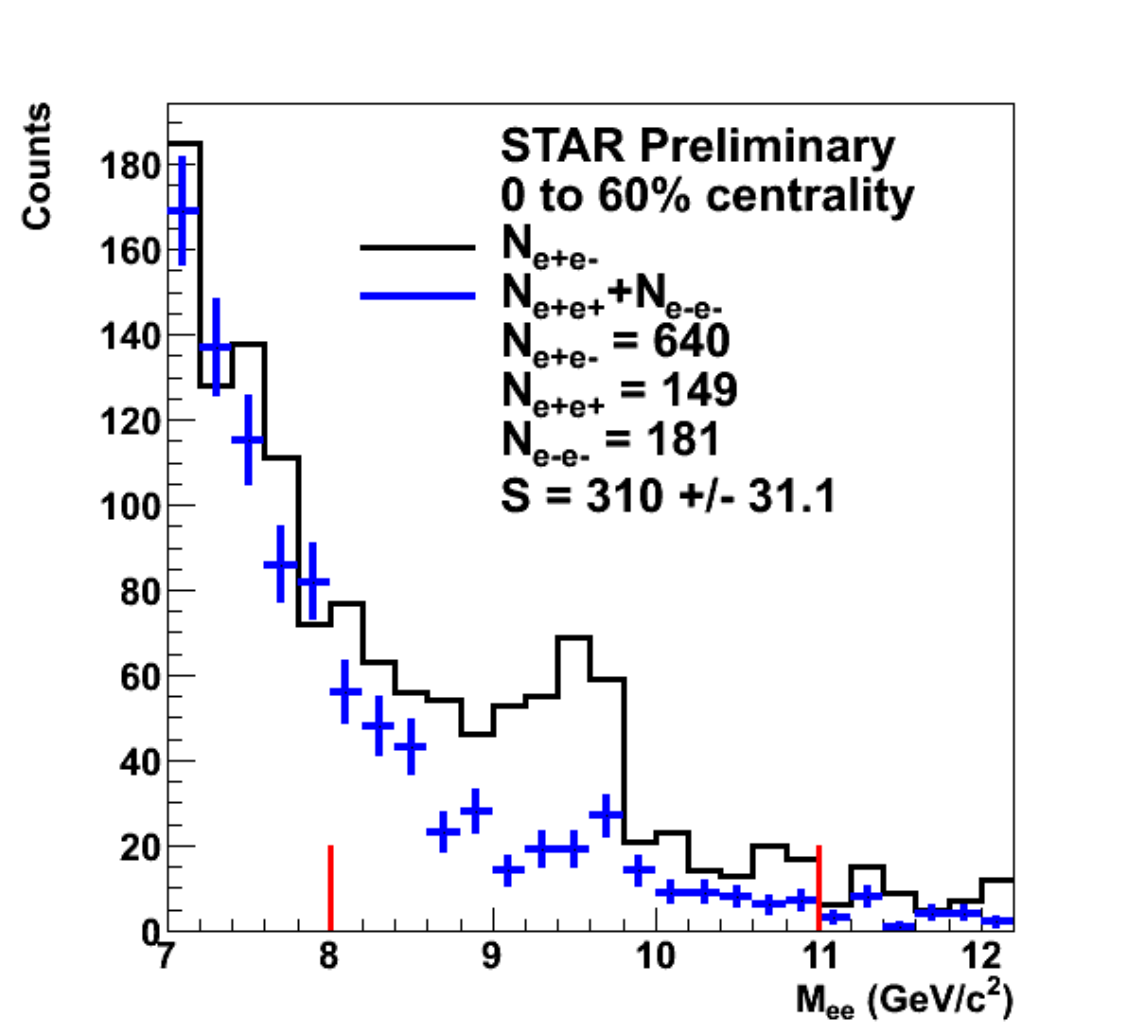}
\includegraphics[width=152px]{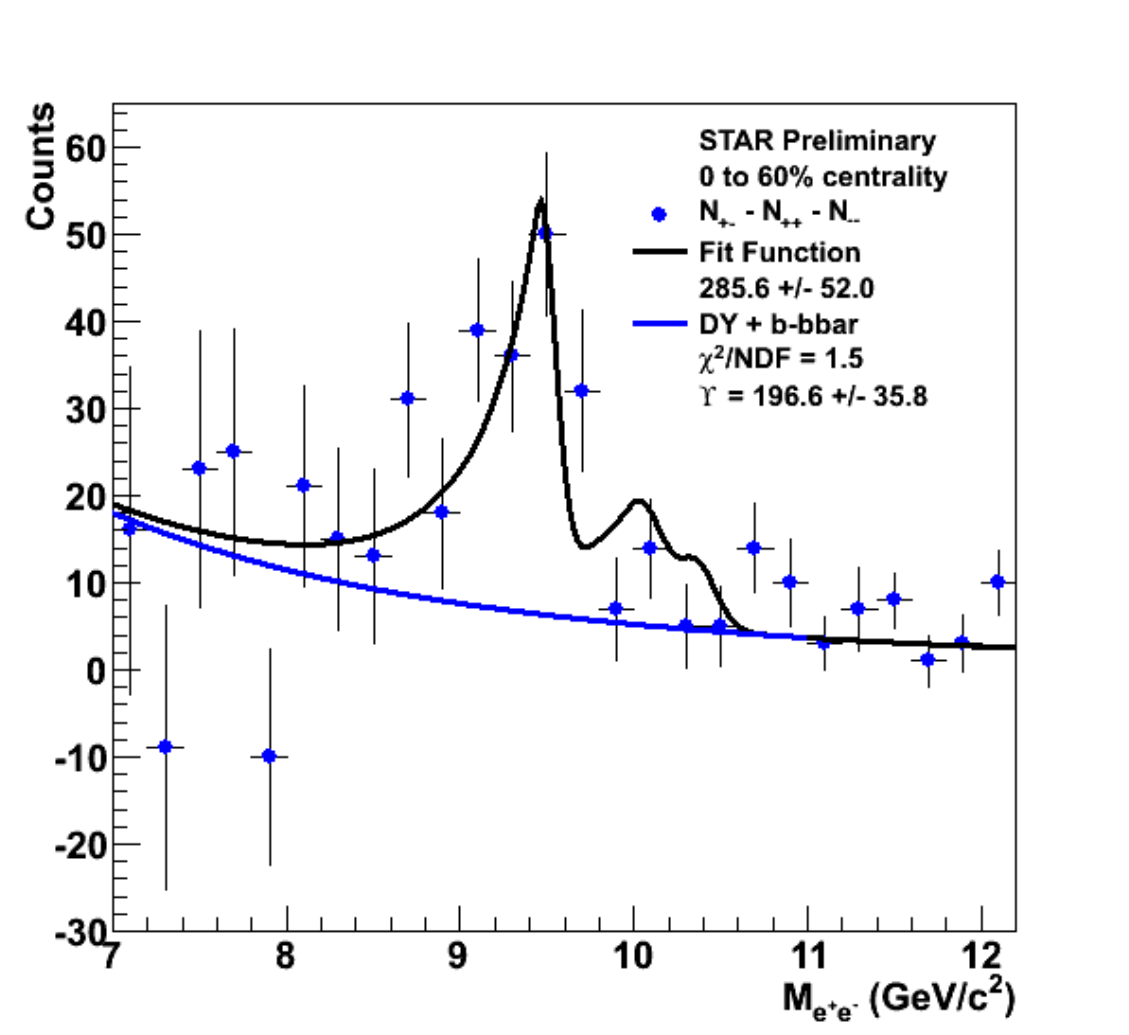}
\includegraphics[width=152px,height=140px,keepaspectratio=false,trim=0 15 0 0]{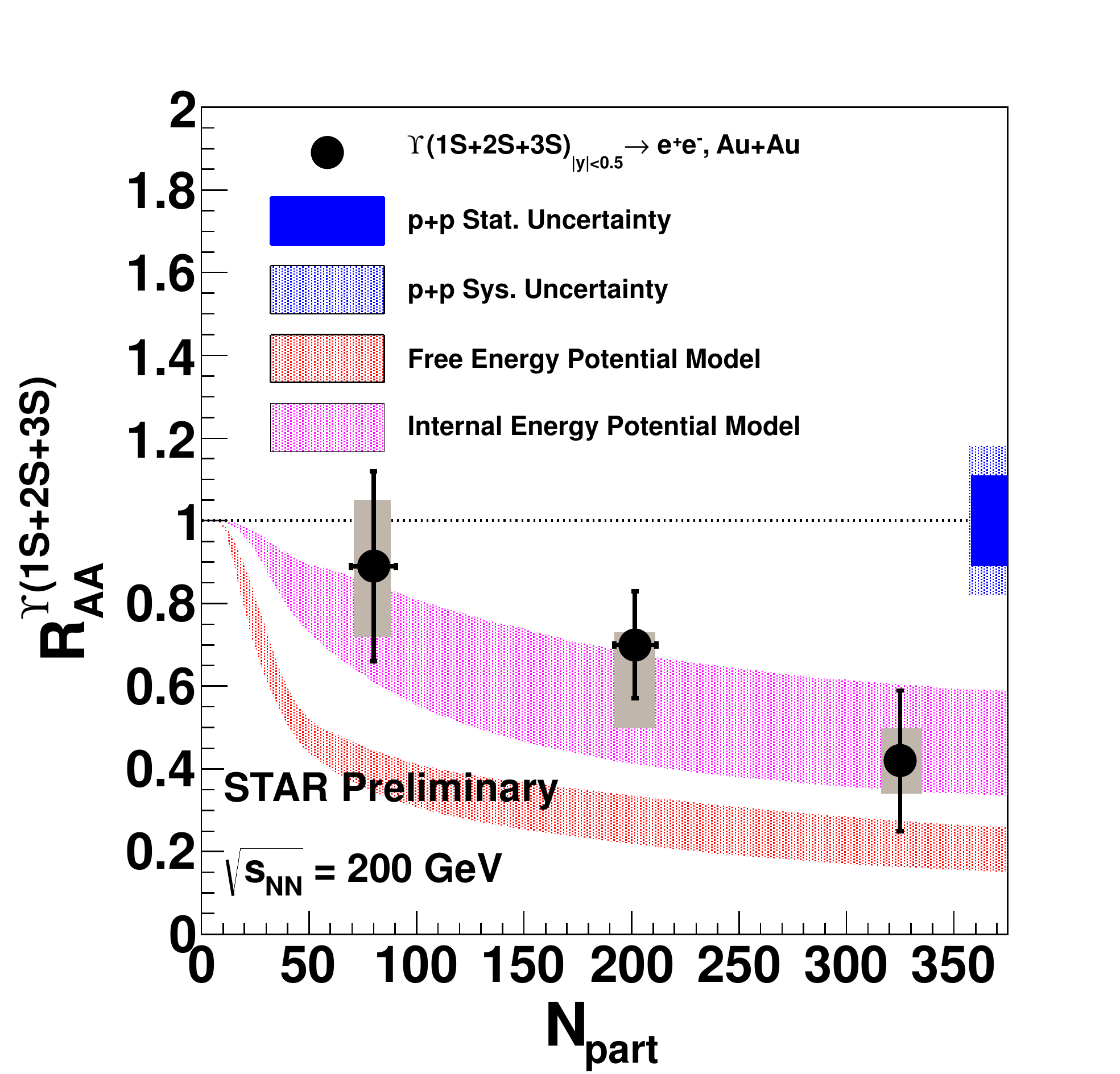}
\caption{\textit{Left:} Invariant mass spectrum of pairs of identified electrons in Au+Au system in the 0\% to 60\% centrality range. Like-sign pairs are in blue; unlike-sign in black. Note that the spectrum no longer has a kinematic peak imposed by the Upsilon trigger. \textit{Center:} Unlike-sign signal subtracted from like-sign signal, 0\%-60\% centrality. The blue curve is the combination of Drell-Yan and \bbbar. The black curve includes the Upsilon signal also. \textit{Right:} \RAA for 2010 Au+Au compared to 2009 p+p. The magenta and orange curves are theoretical predictions using a combination of lattice-based QCD and hydrodynamical expansion and cooling \cite{Strickland:2011kx}.}
\label{fig:AuAu2010_ups}
\end{center}
\end{figure}

Due to an upgrade in our data acquisition system, in 2010 and beyond we could record every event which satisfied the High Tower trigger used to pre-sample the Upsilon trigger. This made the Upsilon trigger obsolete. One advantage of this is that we no longer have a kinematic peak in our background (see Fig. \ref{fig:AuAu2010_ups}, left).

To extract the Upsilon yield, we subtract the like-sign spectrum from the unlike-sign to remove the combinatoric background (see Fig. \ref{fig:AuAu2010_ups}, center). We then fit the remaining signal to a Drell-Yan, \bbbar, and $\Upsilon$(1S+2S+3S) signal. We then remove the fit to Drell-Yan and \bbbar from the subtracted spectrum. Since the lineshape is not used to directly extract the Upsilon yield, this reduces the sensitivity of our measurement to our assumed lineshape or differential state suppression. We find 197$\pm$36 Upsilons in the mass range 8 GeV $<$ m $<$ 11 GeV.

We can separate our measurement into three centrality bins: 0\% to 10\%, 10\% to 30\%, and 30\% to 60\%. Using our 2009 p+p measurement as the baseline, we calculate \RAA for the three centrality bins (see Fig. \ref{fig:AuAu2010_ups}, right). The results are compared to a model by Strickland et al. \cite{Strickland:2011kx} which incorporates lattice-based QCD calculations with a hydrodynamic model of expansion and cooling. Two different forms of the potential are shown. Our data show better agreement with the Internal Energy-based potential. In this model, at central collisions, the 3S is almost completely suppressed, $R_{AA}(2S) \approx 0.2$, and $R_{AA}(1S) \approx 0.4-0.7$. The incomplete suppression is due to differential temperatures in the plasma between the hot core and cooler corona.

\section{Conclusions}
Thanks to the large acceptance of the STAR detector and our dedicated Upsilon trigger, we were able to measure $\Upsilon$ mesons in p+p, d+Au, and Au+Au collisions at $\sqrt{s_{NN}} =200$ GeV. Enhanced statistics from our 2009 p+p run allowed us to refine our measurement of $\Upsilon$ production and reduce the statistical errors by almost a factor of 2. Measured p+p production is consistent with NLO calculations. Our results shows increasing suppression with the number of participants. Furthermore, the results are consistent with hydrodynamical suppression models for the QGP. Future studies will allow us to refine our Au+Au measurements with increased statistics to better understand suppression in the QGP.

This material is based upon work supported by the National Science Foundation under Grant No. 0645773.

%% The Appendices part is started with the command \appendix;
%% appendix sections are then done as normal sections
%% \appendix

%% \section{}
%% \label{}

%% References
%%
%% Following citation commands can be used in the body text:
%% Usage of \cite is as follows:
%%   \cite{key}         ==>>  [#]
%%   \cite[chap. 2]{key} ==>> [#, chap. 2]
%%

%% References with BibTeX database:

\bibliographystyle{elsarticle-num}
\bibliography{hp}

%% Authors are advised to use a BibTeX database file for their reference list.
%% The provided style file elsarticle-num.bst formats references in the required Procedia style

%% For references without a BibTeX database:

% \begin{thebibliography}{00}

%% \bibitem must have the following form:
%%   \bibitem{key}...
%%

% \bibitem{}

% \end{thebibliography}

\end{document}